\documentclass[structabstract]{aa}  
\usepackage{amsmath}
\usepackage{graphicx}
\usepackage{txfonts}
\usepackage[authoryear]{natbib}
\usepackage{booktabs}
\usepackage{lscape}
\bibpunct{(}{)}{;}{a}{}{,}
\usepackage{ marvosym }
\begin{document}

\title{Periodic morphological changes in the \\radio structure of the gamma-ray binary \object{LS~5039}}

\author{J. Mold\'on\inst{\ref{inst1}}\and
M. Rib\'o\inst{\ref{inst1}}\and
J.M. Paredes\inst{\ref{inst1}}}

\institute{$^{1}$Departament d'Astronomia i Meteorologia, Institut de Ci\`encies del
Cosmos (ICC), Universitat de Barcelona (IEEC-UB), Mart\'{\i} i Franqu\`es 1,
08028 Barcelona, Spain\label{inst1}\\ \email{jmoldon@am.ub.es; mribo@am.ub.es; jmparedes@ub.edu}}

\authorrunning{Mold\'on et~al.}
\titlerunning{Periodic morphological changes in the radio structure of \object{LS~5039}}

\date{Received / Accepted}

\abstract
{Gamma-ray binaries allow us to study physical processes such as particle acceleration up to TeV energies and very-high-energy gamma-ray emission and absorption with changing geometrical configurations on a periodic basis. These sources produce outflows of radio-emitting particles whose structure can be imaged with very long baseline interferometry (VLBI). \object{LS~5039} is a gamma-ray binary that has shown variable VLBI structures in the past.}
{We aim to characterise the radio morphological changes of \object{LS~5039} and discriminate if they are either repeatable or erratic.}
{We observed \object{LS~5039} with the VLBA at 5~GHz during five consecutive days to cover the 3.9-day orbit and an extra day to disentangle between orbital or secular variability. We also compiled the available high-resolution radio observations of the source to study its morphological variability at different orbital phases. We used a simple model to interpret the obtained images.}
{The new observations show that the morphology of \object{LS~5039} up to projected distances of 10~milliarcseconds changes in 24~h. The observed radio morphological changes display a periodic orbital modulation. Multifrequency and multiepoch VLBI observations confirm that the morphological periodicity is stable on timescales of years. Using a simple model we show that the observed behaviour is compatible with the presence of a young non-accreting pulsar with an outflow behind it. The morphology is reproduced for inclinations of the orbit of 60--75$^{\circ}$. For masses of the companion star in the range 20--50~M$_{\odot}$, this range of inclinations implies a mass of the compact object of 1.3--2.7~M$_{\odot}$.}
{The periodic orbital modulation of the radio morphology of \object{LS~5039}, stable over several years, suggests that all gamma-ray binaries are expected to show a similar behaviour. The changes in the radio structure of \object{LS~5039} are compatible with the presence of a young non-accreting neutron star, which suggests that the known gamma-ray binaries contain young pulsars.}


\maketitle

\section{Introduction} \label{introduction}

Gamma-ray binaries are extreme systems that produce non-thermal emission from radio to very-high-energy ($>$TeV) gamma rays, with the energy output in the spectral energy distribution (SED) dominated by the MeV--GeV photons (see \citealt{paredes11} for a review). Their broadband emission is usually modulated by the orbital cycle of the system, which suggests that the physical conditions are also periodic and reproducible. The wide range of different orbital periods and eccentricities of the known gamma-ray binaries \cite[see Table~4 in][]{casares12hess}, provides a diversity of different ambient conditions in which the physical processes take place. The diversity of systems, together with the reproducibility of the conditions within each system, makes gamma-ray binaries excellent physical laboratories in which high energy particle acceleration, diffusion, absorption, and radiation mechanisms can be explored. Nevertheless, the number of known gamma-ray binaries is still very limited. Five binary systems have been classified as gamma-ray binaries and present orbital modulation of the GeV and/or TeV gamma-ray emission: \object{PSR~B1259$-$63} \citep{aharonian09_psr}, \object{LS~5039} \citep{aharonian06, abdo09_ls}, \object{LS~I~+61~303} \citep{albert09,abdo09_lsi}, \object{HESS~J0632+057} \citep{aharonian07, maier12}, and \object{1FGL~J1018.6$-$5856} \citep{ackermann12}. The latter has not been clearly detected at TeV energies yet. On the other hand, \object{Cygnus~X-3} presents active periods of GeV emission in which orbital periodicity is detected \citep{abdo09_cyg_x3,tavani09}. \object{Cygnus~X-1} showed evidence of gamma-ray emission at TeV energies on a single night of observations \citep{albert07} . However, in these two cases the peak of the SED is at keV energies and therefore are classified as X-ray binaries, the high energy activity of which is clearly powered by accretion. Another phenomenological difference between X-ray and gamma-ray binaries is that variations in the non-thermal broadband emission in X-ray binaries is usually linked to X-ray state changes, whereas in gamma-ray binaries it is coupled to the orbital period.

\object{LS~5039} is a binary system composed by the homonym bright star \object{LS~5039}, of spectral type ON6.5\,V((f)), and a compact object \citep{casares05}. The orbital period of the system is 3.9~days and the eccentricity of the orbit is $e\sim0.35$ \citep{casares05,aragona09,sarty11}. Given the mass function of the binary system the compact object would be a black hole for low inclinations of the orbit, and a neutron star for high inclinations (see the discussion in \citealt{casares05} and recent constraints in \citealt{casares12}). No short-period pulsations were found that could demonstrate the presence of a pulsar either in radio \citep{mcswain11} or X-rays \citep{martocchia05,rea11}. The distance to the system has recently been updated to $2.9\pm0.8$~kpc \citep{casares12}. \object{LS~5039} displays non-thermal, persistent, and periodic gamma-ray emission up to 4~TeV \citep{paredes00,aharonian05,abdo09_ls}. The X-ray and gamma-ray emission show a periodic modulation of 3.9~days \citep{bosch-ramon05, aharonian06, kishishita09, abdo09_ls}. The total radio emission of \object{LS~5039} is variable, although no periodicity or strong outbursts have been detected so far \citep{marti98, ribo99, ribo02_phd}. The radio emission above 1~GHz is non-thermal, with spectral index of $-0.46$ \citep{marti98}, and inverted at lower frequencies (\citealt{godambe08, bhattacharyya12}, but see \citealt{pandey07}). The synchrotron radio emission appears extended when observed with VLBI on scales of milliarcseconds (mas). The source shows a main core and extended bipolar emission that has been observed in directions with position angles (P.A.) between 120 and 150$^{\circ}$ and projected angular distance between 1 and 180~mas (3--500~AU) from the core \citep{paredes00,paredes02,ribo08}. In particular, \cite{ribo08} measured a change in the source morphology at scales of 2--6~mas from VLBI images obtained 5~days apart. Finally, for X-rays, \cite{durant11} discovered an extended component in X-rays up to 2$^{\prime}$ from \object{LS~5039}.

Other gamma-ray binaries also present extended and variable radio emission at mas scales. \object{PSR~B1259$-$63} is the only gamma-ray binary in which radio pulsations have been detected up to now. During the periastron passage of this eccentric system, occurring every 3.4~yr, the close interaction between the pulsar and the massive Be star and/or Be disc produces broadband emission during a few months. \cite{moldon11_psr} discovered transient extended emission during and after the periastron passage with a total projected extent of $\sim50$~mas (120~AU). The peak of the radio emission was detected outside the binary system. VLBI observations at different orbital phases of the 26.5-d period binary \object{LS~I~+61~303}, which also hosts a Be star, were conducted by \cite{dhawan06}, showing an orbital variability that these authors interpreted as the signature of a cometary tail produced in the colliding winds scenario. Another interpretation of the data suggests that the changes are compatible with a precessing microblazar \citep{massi12}. The system \object{HESS~J0632+057}, also hosting a Be star, has an orbital period of 321~days \citep{bongiorno11}, and faint radio extended emission at mas scales was detected $\sim130$~days after the periastron passage \citep{moldon11_hess}. For \object{1FGL~J1018.6$-$5856}, with a period of 16~d and also hosting an O6V((f)) star \citep{ackermann12}, no VLBI observations have been reported so far. In addition, some of these systems also present extended X-ray emission at larger scales of $\sim1^{\prime}$, as hinted for \object{LS~I~+61~303} \citep{paredes07} and shown for \object{PSR~B1259$-$63} \citep{pavlov11}.

Several theoretical models were developed to explain the multiwavelength behaviour of \object{LS~5039}. The very-high-energy gamma-ray emission can be interpreted as the result of inverse Compton upscattering of stellar UV photons by relativistic electrons. The acceleration of electrons can be explained by two exclusive scenarios: acceleration in the jet of a microquasar powered by accretion (\citealt{paredes06, bosch-ramon06} and the review in \citealt{bosch-ramon09}), or shocks between the relativistic wind of a young non-accreting pulsar and the wind of the stellar companion \citep{maraschi81, tavani97, dubus06,khangulyan07}. A simple and shockless microquasar scenario is disfavoured by previous VLBI radio observations of \object{LS~5039} \citep{ribo08}. Also, no signs of the presence of an accretion disc have been detected so far (although see \citealt{barkov12} and \citealt{okazaki08} for alternative explanations). \cite{sarty11} used the stability of the optical photometry of \object{LS~5039} to constrain the orbit inclination to be below 60$^{\circ}$ and the mass of the compact object above 1.8~M$_{\odot}$. Models based on the young non-accreting pulsar scenario, first proposed for \object{LS~5039} in \cite{martocchia05}, had provided descriptions of the HE/VHE light curves and the spectral evolution of the source as a function of the orbital phase \citep[see][]{sierpowska-bartosik07, dubus08, khangulyan08, takahashi09}. Apart from the acceleration of particles in the main shock between winds, additional gamma-ray emission can be produced by the unshocked pulsar wind \citep{cerutti08} and by secondary cascading \citep{bosch-ramon08, cerutti10}. The X-ray light curve and spectrum show orbital variability, although there are no signatures of variable X-ray absorption or X-ray occultations, as discussed in \citet{reig03}, \citet{martocchia05}, \citet{bosch-ramon07} and \citet{szostek11}. \cite{zabalza11} constrained the spin-down luminosity of the putative pulsar to be $\sim3$--$6\times10^{36}$~erg~s$^{-1}$.

Very recently, detailed hydrodynamical simulation have been obtained to understand the wind shocks and the flow dynamics in gamma-ray binaries. Two-dimensional hydrodynamic simulations were performed to study the wind-wind collision interaction at scales of the binary system for \object{PSR~B1259$-$63} \citep{bogovalov08, bogovalov12} and for \object{LS~I~+61~303} \citep{romero07}. Also for \object{PSR~B1259$-$63}, \cite{okazaki11} and \cite{takata12} presented 3D simulations of the tidal and wind interactions. \cite{lamberts12} considered the general case of colliding wind binaries and described the outflow structure at larger distances. Finally, the flow structure for a case similar to \object{LS~5039} was described in \cite{bosch-ramon12} for scales up to 100~times the orbit size.

From a theoretical point of view, a general discussion on the properties of the radio emission of \object{LS~5039} can be found in \cite{bosch-ramon08}. However, only a few predictions on the expected radio morphology have been discussed. \cite{dubus06} presented emission maps at different orbital phases for \object{LS~5039} and predicted orbital morphological changes at mas scales, as well as displacements of the peak of the emission of a few mas. \cite{bosch-ramon11_second} obtained synthetic maps at radio wavelengths from a model based on secondary particles created from gamma-ray absorption, which may account for a significant fraction of the total radio flux density of the source and also predicts extended radio emission at mas scales.

In this paper we present the results from a dedicated VLBA campaign that covers a full orbital cycle and provides astrometric and morphological information during five consecutive days. The observations and data reduction are presented in Sects.~\ref{observations} and \ref{reduction}, including a discussion on the calibration caveats. In Sect.~\ref{results} we present the obtained astrometry and morphology of the source at different orbital phases and analyse the components characterising the extended emission of the source. In Sect.~\ref{compilation} we show a compilation of VLBI data on \object{LS~5039} and we describe and compare the morphology at different orbital phases for observations at frequencies between 1.7 and 8.5~GHz taken from 1999 to 2009. In Sect.~\ref{model} we present a model to check if the measured changes are compatible with a wind-wind collision scenario. The model allows us to estimate the orientation of the orbit on the sky, in particular its inclination, and consequently constrain the mass of the compact object. Finally, we present a summary of the results and the main conclusions of this work in Sect.~\ref{conclusions}.

\section{Observations} \label{observations}

We conducted VLBI observations on \object{LS~5039} during five consecutive days to cover an orbit of 3.9~d and an extra day to disentangle between orbital or secular variability. We observed at 5~GHz (6~cm wavelength) with the Very Long Baseline Array (VLBA) of the National Radio Astronomy Observatory (NRAO). The stations forming the array are Br, Fd, Hn, Kp, La, Mk, Nl, Ov, Pt, and Sc. The VLBA project code is BR127, and the five observations were conducted from July 5 to 9, 2007 (MJD~54286 to 54290). The five runs, hereafter run A--E, spanned from 03:30 to 09:30~UTC and were scheduled in the same way to reduce differences between runs. The corresponding orbital phases were computed using the ephemerides in \cite{casares12}, $T_{0}={\rm HJD~}2\,453\,478.09(6)$ and $P_{\rm orb}=3.90603(8)$, where the values in parentheses refer to the uncertainty in the last digit. The phase of the periastron passage is 0.0. The orbital phase at the centre of each observation was 0.03, 0.29, 0.55, 0.80, and 0.06, respectively, with an uncertainty of 0.02, or approximately 1.9~hours. A log of the observations is shown in Table~\ref{table:obsparam}.

\begin{table} 
\begin{center}
\caption{Log of VLBA observations of project BR127, conducted at 5~GHz. The orbital phases have an uncertainty of 0.02.}
\label{table:obsparam} 
\begin{tabular}{c cc c }
\midrule
\midrule
Run &   Date (Y-M-D) & MJD                  & Phase range        \\
\midrule
A   &   2007-07-05   &  54286.15--54286.40   &  0.00--0.07         \\
B   &   2007-07-06   &  54287.15--54287.40   &  0.26--0.32         \\
C   &   2007-07-07   &  54288.15--54288.40   &  0.51--0.58         \\
D   &   2007-07-08   &  54289.15--54289.40   &  0.77--0.83         \\
E   &   2007-07-09   &  54290.15--54290.40   &  0.03--0.09         \\
\midrule
\end{tabular}
\end{center}
\end{table}

The data were obtained with single circular left-handed polarisation, with eight sub-bands of 8~MHz, and were correlated with two bits per sample, obtaining a total aggregate bit-rate of 256~Mbps. The data were processed at the VLBA hardware correlator in Socorro, which produced the final visibilities with an integration time of 2 seconds.

\begin{figure}[h] 
\center
\resizebox{1.0\hsize}{!}{\includegraphics[angle=0]{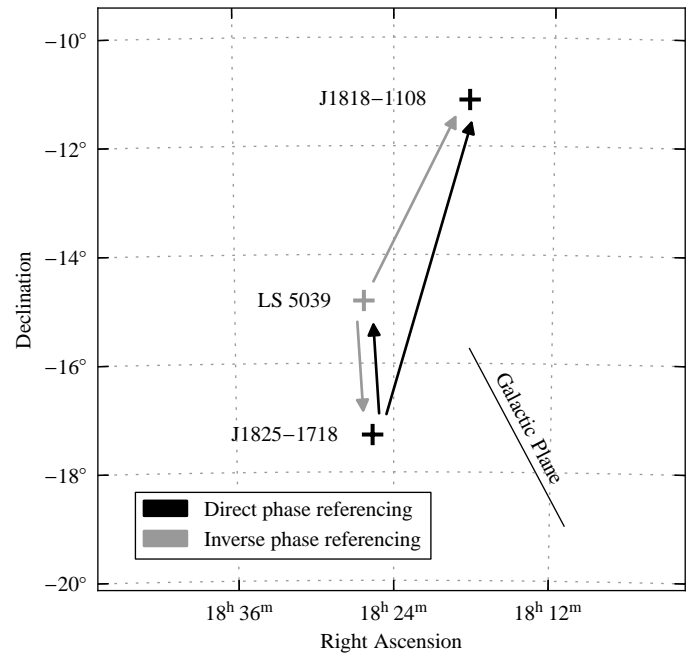}}
\caption{Distribution on the sky of \object{LS~5039}, the phase calibrator \object{J1825$-$1718}, and the astrometric check source \object{J1818$-$1108}. The dashed line indicates the Galactic Plane. The arrows indicate the two phase calibrations obtained: direct, using \object{J1825$-$1718} as phase reference; inverse, using \object{LS~5039} as phase reference.}
\label{fig:sky}
\end{figure}

We conducted the observations using the phase-referencing technique on the phase reference calibrator \object{J1825$-$1718}, located at 2.5$^{\circ}$ from \object{LS~5039} in a P.A. of $-176.4^{\circ}$ (see Fig.~\ref{fig:sky}). We observed in 4.2-min cycles, spending 2.5~minutes on the target source and 1.7~minutes on \object{J1825$-$1718}. The total scheduled time on the target source was 2.7~hours. As an astrometric check source we used \object{J1818$-$1108}, including one scan of 2.5~minutes every 30~minutes. Finally, we included three 5-minute scans on the fringe finder J1733$-$1304. The reference position for the observations is the correlation position of the phase reference source \object{J1825$-$1718} $\alpha_{\rm J2000.0}=18^{\rm h} 25^{\rm m} 36\fs53228\pm0\fs00009$ (or $\pm1.3$~mas) and $\delta_{\rm J2000.0}=-17\degr 18\arcmin 49\farcs848\pm0\farcs002$ (or $\pm2$~mas), taken from \cite{fomalont03} via the SCHED database (NASA catalogue 2005f\_astro). \object{LS~5039} is close to the galactic plane, at a galactic latitude of $-$1.29$^\circ$. The phase calibrator and the astrometric check source suffer from galactic scatter broadening \citep[a general discussion on this effect can be found in][]{fey91}. The phase calibrator \object{J1825$-$1718} has a total flux density of 350~mJy at 5~GHz, but the amplitude of the visibilities decreases with the $uv$ distance and the phase calibration fails for baselines longer than $\sim$90--100~M$\lambda$. The compact core has a size of 3~mas. The resolution of the phase-referenced images is therefore limited by the scatter broadening of the phase calibrator. On the other hand, the astrometric check source \object{J1818$-$1108} is not compact. The self-calibrated images show a main core of 680~mJy and a secondary blob of 180~mJy located at 50~mas westwards from the core. Both components have a size of $\sim$8~mas. The visibility amplitudes quickly drop with the $uv$ distance and there is no signal beyond $\sim$30~M$\lambda$. The distance between \object{J1818$-$1108} and the phase calibrator is 6.4$^\circ$ in P.A. of 16$^\circ$ (see Fig.~\ref{fig:sky}), which makes difficult to transfer the phase solutions for this low elevation source.

\section{Data reduction} \label{reduction}

Data reduction was principally performed in {\sc AIPS}\footnote{The NRAO Astronomical Image Processing System. http://www.aips.nrao.edu/}. The Difmap package \citep{shepherd97} was used for imaging and self-calibration. The raw visibilities were loaded into AIPS, and a priori flagging on telescope off-source times because of antenna slewing was applied. We searched for data with instrumental problems and flagged them, as well as all visibilities with antenna elevation below 5$^{\circ}$. Initially, we updated the geometrical model of the correlator using the Earth orientation parameters (EOPs). However, the correction implied a change on the final measured positions one order of magnitude below our final uncertainties. Considering the small effect and the problems with this correction in the past (wrong parameters during the correlations between 2003 and 2005\footnote{http://www.vlba.nrao.edu/memos/test/test69memo/eop\_problem.html } and a bug in the AIPS task CLCOR from 2009 to 2011), we did not include them in the final data processing. For these observations, the EOPs correction produced a $\sim10^{\circ}$ phase offset in the visibilities, which was in any case removed by the phase calibration (see below).

\subsection{Ionospheric correction} \label{ionospheric}

The different ionospheric conditions above VLBI antennas are one of the main contributions to systematic errors for astrometry at low frequencies. The unmodelled phase and delay contribution of the ionosphere modifies randomly the observed phases on timescales of minutes, producing two effects. On one hand, it spreads the signal and broadens the source, producing an overall decrease of the detected signal-to-noise ratio (S/N). On the other hand, it introduces an unknown position offset. This correction is specially important in this case because the sources have low declination, which implies low elevations during the observations, and because the sources are separated mainly in the North-South direction (see Fig.~\ref{fig:sky}), where the atmospheric conditions change more significantly.

We used ionospheric total electron content (TEC) models based on GPS data obtained from the CDDIS data archive\footnote{The Crustal Dynamics Data Information System http://cddis.nasa.gov/} to correct the phase variations caused by the ionosphere. These maps provide one TEC measures every two hours on a grid of $5^{\circ}\times2.5^{\circ}$ in terrestrial longitude and latitude, respectively. This is a coarse correction because the ionosphere changes occur on a shorter timescale, although it accounts for significant  astrometric offsets. The ionospheric models are produced by different institutes, which estimate the TEC above different locations and provide electron distribution maps (IONEX files) that can be loaded into AIPS to apply the phase corrections. We compared the effects of the phase correction on the visibilities using the models provided by four institutes: the Jet Propulsion Laboratory (JPL), the Center for Orbit Determination in Europe (CODE), the ESOC Ionosphere Monitoring Facility (ESA), and the Universitat Polit\`{e}cnica de Catalunya (UPC). We tested the models by comparing the S/N and the stability of the position of the peak of the emission of \object{LS~5039} in different short intervals of 1 and 3 hours along the observations. We did not measure appreciable differences in the peak flux density when using the different models. The ionospheric models provide an average position offset of 0.02 and $-1.7$~mas, in right ascension and declination, respectively. The models are very similar, and the dispersion of these corrections are 0.11 and 0.2~mas, respectively. For each run we applied the model that provides offsets closer to the averaged value. The JPL model was used to correct the runs A and C, and the CODE model to correct runs B, D, and E.

We checked the stability of the correction for the observed times by computing the variability of the TEC provided by the GPS maps. The most variable stations were Sc (at the end of the observation the Sun was already rising) and Mk (the observations started before sunset). Comparing the mean variability of the TEC content for every run during the central hours of the project (from 4:00 to 8:00 UTC), we found that runs B, D, and E have a relative average variability of 18\%, for run~C it was 15\%, and for run~A 22\%. This shows that during run~A the ionosphere was, in general, more variable. This can in part justify the astrometric problems encountered for this run (see Sect.~\ref{phasereferencing}), being the rest of the weather conditions similar for all runs.

\subsection{Amplitude and phase calibration} \label{calibration}

ACCOR was used to fix the amplitudes in the cross correlation spectrum from the VLBA correlator. The amplitude calibration was performed using the antenna gains and the system temperatures measured at each station in real time during the observations, using the APCAL procedure. The parallactic angle correction was automatically applied with VLBAPANG. To correct the instrumental offsets and slopes between and within the different bands we ran FRING on 30 seconds of a scan of the fringe finder \object{3C~345} (manual phase-cal). To correct the dependence of the visibility amplitudes with the frequency we used the auto-correlation values from \object{3C~345} to smooth the bandpass shape of the amplitudes.

An initial fringe fitting was performed using the FRING routine in AIPS, using a point-source model on the phase calibrator \object{J1825$-$1718}. These data were exported to Difmap and self-calibrated. We fitted the visibilities for each run with a circular Gaussian. The mean flux density fitted for the core component was 345~mJy, with a standard deviation between runs of 2.3~mJy (0.67\%), while the calibrator structure did not change significantly. Therefore, we combined the five data sets of the calibrator and performed several iterations of imaging and self-calibration to create a master calibrator image. We ran FRING again, but now we removed the source structure contribution to the phase calibration by using the combined image of \object{J1825$-$1718} as an input model. One solution for each scan for all IFs was searched within a delay and rate search windows of 80~nanoseconds and 20~mHz, respectively. The minimum allowed S/N for solutions was 8, providing 95.8\% of good solutions. The 4.2\% of the data without solutions mostly correspond to long baselines (above 90~M$\lambda$). The solutions for every frequency band were corrected with MBDLY, which fits the phase delay between bands. The final phases were explored and flagged when necessary. Finally, we applied the calibration and flag tables, and we averaged all frequency channels within each IF to obtain three single-source files, one for \object{LS~5039}, one for the phase calibrator \object{J1825$-$1718} and one for the astrometric check source \object{J1818$-$1108}.

We also obtained inverse phase referencing based on \object{LS~5039} because the source can be self-calibrated at 5~GHz. We followed an analogue procedure to fringe-fit the \object{LS~5039} data, although we did not produce a combined data set. The S/N cut was set to 5 because of the faintness of the source. We flagged the unstable phases and transferred the solutions from \object{LS~5039} to \object{J1825$-$1718} and \object{J1818$-$1108}, for which inverse astrometry was obtained. A sketch of the phase referencing schemes is shown in Fig.~\ref{fig:sky}.

\subsection{Phase referencing imaging} \label{phasereferencing}

To reduce the systematic errors of the phase referencing produced by data taken when the sources were at low elevations, we only used the central hours of the runs. We produced images with time intervals of 30 and 60~minutes and inspected the image quality and the astrometric stability. We also produced images for the individual scans. The data obtained at the beginning and at the end of the observations are not consistent because we measured artificial displacements of the peak of the emission between $\sim$0.5 and 3~mas, while the source became elongated in random directions. This behaviour can be explained by the ionosphere affecting the data with low elevation. We note that the effect is more important for run~A, which suffered severe displacements of more than 6~mas in the time blocks for the first hour and the last two hours. After inspecting these systematic errors for different time blocks we only used the data in the time intervals when the source was reasonably stable, from 04:40 to 07:40 UT for runs B, C, and D, from 04:30 to 07:00 UT for run~A, and from 06:00 to 08:30 UT for run~E. Taking into account that the declination of the phase-reference calibrator and the target source are $-17.3^{\circ}$ and $-14.8^{\circ}$, respectively, that the observations were centred at the culmination of the sources, and that the observations were conducted under normal weather conditions (without rain, strong winds or snow), the time ranges used correspond to elevations of most of the antennas above $\sim25^{\circ}$. Some antennas contributing to the longest baselines (Mk, Sc, Hn, and Br) were below this value during part of the observations, and the phase calibration was not succesful for them during these time intervals, altough this can be also caused by the intrinsic calibrator structure.

The phase-referenced data were imaged using task IMAGR with a pixel size of 0.2~mas. We used a weighting scheme with robust~0 as a compromise between angular resolution and sensitivity. \cite{pradel06} showed that removing the shorter baselines of the VLBA can decrease the astrometric systematic errors by a 15\%. For \object{LS~5039}, the source was detected with S/N above 10 when imaging the baselines with $uv$ distances between 15--20 and 100~M$\lambda$. The range of minimum baselines was lowered in those cases in which secondary lobes were more important or when the S/N of the peak emission was too low. The same strategy was used to image the astrometric check source \object{J1818$-$1108}. In this highly resolved source we only used baselines with $uv$ distances between 12 and 40~M$\lambda$. 

For the inverse phase referencing, where phases from \object{LS~5039} were transferred to \object{J1825$-$1718} and \object{J1818$-$1108}, a $uv$ range of 20 to 60~M$\lambda$ was used for \object{J1825$-$1718} and 12 to 60~M$\lambda$ for the check source. The maximum baseline of the latter is higher than in the direct phase referencing because of the smaller angular distance between \object{J1818$-$1108} and the phase reference source in this case (see Fig.~\ref{fig:sky}). In both cases we used a pixel size of 0.5~mas.

\subsection{Self-calibration} \label{selfcalibration}

In order to perform the self-calibration of the \object{LS~5039} data, we loaded the individual data files into Difmap. We averaged the data with a binning size of 50~seconds. The visibilities with long baselines were down-weighted, using a Gaussian taper at radius of 80~M$\lambda$. We performed several iterations of cleaning and phase self-calibration. For each of them, phase solutions were found, first for the  most compact part of the core by imaging the data with a uniform weighting scheme and then solving for the more extended emission using a natural weighting cleaning. After each cycle, an amplitude self-calibration was obtained, each time with a shorter integration time, up to a minimum of 30~minutes. Finally, two hybrid images were produced for each run, one with natural weight (worse resolution and better sensitivity) and one with uniform weight (better resolution and worse sensitivity). The final images were exported to AIPS. Their rms noise and the total flux density were obtained by fitting the pixel flux density distribution of the image with IMEAN. The extended emission was characterised by fitting two or three Gaussian components to the images using the task JMFIT.

We also fitted Gaussian components to the calibrated $uv$ data sets of \object{LS~5039} (modelfit). This is an iterative process, where the initial values for the fit have to be fixed manually, and the final solution can depend on the starting parameters. Therefore, we note that the model fitting results can be slightly subjective, in particular for this case, where the source structure is not clearly defined by individual components. Despite this caveat, the modelfit provides morphological information directly from the $uv$ data and therefore it is not limited by a synthesized beam because it is independent of the deconvolution procedure used. We obtained the modelfit of the $uv$ data with the task UVFIT in AIPS. As a starting point, we used the solutions from JMFIT obtained from the naturally weighted images. Then we inspected several combinations of number of components and shape restrictions until a robust solution was found. As a parallel check, we performed another modelfit in Difmap to verify the consistency of the results. As both approaches provide similar results, we only present here the solutions obtained with AIPS.

\section{Analysis and results} \label{results}

\subsection{Astrometry} \label{astrometry}

The average position of \object{LS~5039} from the five runs of project BR127 is $\alpha_{\rm J2000.0}=18^{\rm h} 26^{\rm m} 15\fs06003(3)$ (or $\pm0.4$~mas) and $\delta_{\rm J2000.0}=-14\degr 50\arcmin 54\farcs3094(6)$ (or $\pm0.6$~mas). This position is measured with respect to the correlation position of \object{J1825$-$1718}, which is listed in Sect.~\ref{observations}. The absolute astrometry in this project was used to obtain an accurate proper motion of \object{LS~5039} \citep{moldon12astrom}. The astrometric uncertainties cannot be obtained directly from the position fit because of the unknown ionospheric effects, and therefore we used the astrometric check source \object{J1818$-$1108} to estimate them. The standard deviation of the peak positions of the check source was 2.3~mas in right ascension and 2.4~mas in declination. These deviations were converted to uncertainties in the determination of the position of \object{LS~5039} by multiplying them by two correction factors that account for the different observing conditions for \object{LS~5039} and \object{J1818$-$1108}, following the general theoretical astrometric precision for an interferometer \citep{thompson86}

\begin{equation}
\sigma = \frac{1}{2\pi}\frac{1}{S/N}\frac{\lambda}{B},
\end{equation}
\noindent
where the wavelength ($\lambda$) and the maximum baseline ($B$) can be represented by the synthesized beam size. First, the uncertainty was scaled taking into account the different resolutions of the images. Basically, the data of \object{J1818$-$1108} were imaged with poorer resolution, only using short baselines (see Sect.~\ref{phasereferencing}). This factor was obtained as a ratio of the synthesized beam size of the \object{LS~5039} observations to the mean size of the \object{J1818$-$1108} beams. Second, the astrometric precision depends on the S/N of the detection. We scaled the uncertainty by the ratio of individual S/N of \object{LS~5039}, which was around 15, to the mean S/N of \object{J1818$-$1108}, which was 7. These two corrections yield a scaling factor applied to the dispersion of \object{J1818$-$1108} of $\sim0.19$ and $\sim0.25$ in right ascension and declination, respectively. Finally, the mean uncertainties in the determination of the right ascension and declination of \object{LS~5039} are 0.4 and 0.6~mas, respectively. The same procedure was used for the inverted astrometry.

Another approach to estimate the astrometric uncertainties is to use the expected theoretical systematic uncertainties computed using Eq.~2 in \cite{pradel06}, although some assumptions considered in the general discussion in that paper are not met in these observations. The minimum theoretical uncertainties for \object{LS~5039} using \object{J1825$-$1718} as a reference source and assuming mean tropospheric conditions are 0.17 and 0.56~mas in right ascension and declination, respectively, which are compatible with the estimation quoted above. This indicates that the uncertainties in declination seem to be dominated by systematic uncertainties. On the other hand, the minimum theoretical uncertainties for \object{J1818$-$1108} are 0.43 and 1.42~mas in right ascension and declination, respectively. The real dispersion measured, 2.3~mas in right ascension and 2.4~mas in declination, is significantly larger possibly because the extended nature of the source, the poor $uv$ coverage, and the low S/N. Given these results, we have adopted the more conservative approach described above, although we note that the real uncertainties of the measurement are possibly a non-trivial combination of the systematic and nominal uncertainties.

\begin{figure}[] 
\center
\resizebox{1.0\hsize}{!}{\includegraphics[angle=0]{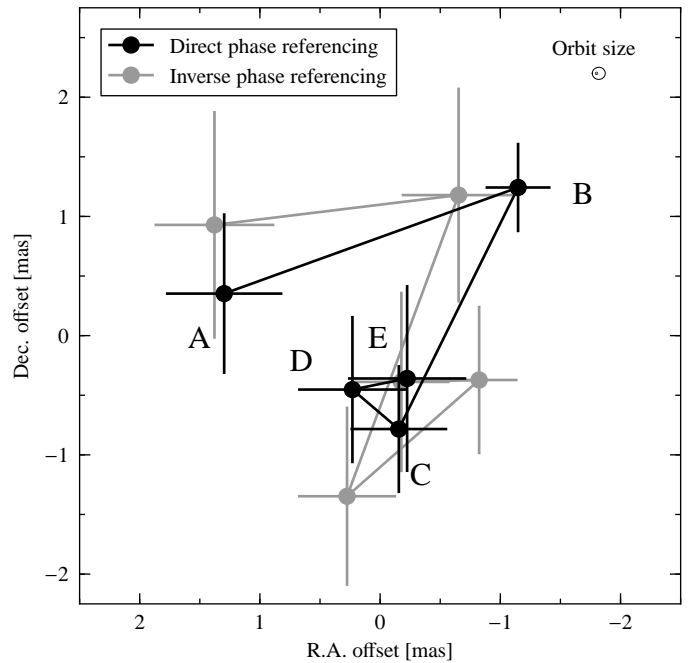}}
\caption{Relative offsets of the peak of the emission of \object{LS~5039} during five consecutive days with respect to the average position. For straightforward comparison, we plot the opposite value of the inverse astrometry. The run label is displayed close to each direct measure. The size of the binary system is indicated by the small orbit in the corner, which is plotted face-on and with an arbitrary orientation. The displacements and uncertainties are shown in Table~\ref{table:astrometry}.}
\label{fig:astrometry}
\end{figure}

In Table~\ref{table:astrometry} we present the measured astrometry. We show the relative astrometry (with respect to the average position) for the check source \object{J1818$-$1108} and the synthesized beam of the phase referenced images and the relative astrometry for \object{LS~5039}. We also show the inverted astrometry obtained using \object{LS~5039} as a phase reference source. The measured relative offsets of the peak of the emission of \object{LS~5039} using the two methods are plotted in Fig.~\ref{fig:astrometry}. Both approaches provide similar relative astrometry. The peak position of \object{LS~5039} during run~A is significantly displaced from the mean value. However, in run~A, the check source shows a even larger displacement in the same direction (see first row in Table~\ref{table:astrometry}). This fact, combined with the suspicious behaviour described in Sect.~\ref{phasereferencing}, makes clear that the astrometry of run~A suffers from uncorrected ionospheric effects, and therefore we do not consider the peak displacement in run~A as reliable. The only significant displacement is found between run~B and run~C, which corresponds to displacements of $1.0\pm0.5$~mas and $-2.0\pm0.7$~mas in right ascension and declination, respectively. Therefore, the total displacement between phase 0.29 and phase 0.55 is $2.3\pm0.6$~mas in P.A. of $154^{\circ}$, which, considering the limitations of the data, cannot be considered a robust sign of peak displacement.

\begin{table*}[ht] 
\begin{center}
\caption{Relative astrometry of the five runs of project BR127. The displacements of the check source J1818$-$1108 are used to determine the astrometric uncertainties (see Sect.~\ref{astrometry}). The mean synthesized beam for J1818$-$1108 in the direct phase referencing is $6.5\times3.3$~mas$^{2}$ at P.A. of $6^{\circ}$, and for the inverse phase referencing it is $6.9\times2.7$~mas$^{2}$ at P.A. of $7^{\circ}$. We show, for \object{LS~5039} and \object{J1825$-$1718}, the beam (HPBW) size and P.A. and the relative displacements measured from the average position. These results are plotted in Fig.~\ref{fig:astrometry}.}
\label{table:astrometry} 
\begin{tabular*}{\textwidth}{@{\extracolsep{-0.25pt}}c  c c cc r@{ $\pm~$ }l r@{ $\pm~$ }l c c c cc r@{ $\pm~$ }l r@{ $\pm~$ }l @{}}
\midrule
\midrule
         &  \multicolumn{8}{c}{Phase referencing from J1825$-$1718 (direct)}  &  \multicolumn{8}{c}{Phase referencing from LS~5039 (inverse)}  \\
\cmidrule{2-9} \cmidrule{11-18}
Run      &  \multicolumn{2}{c}{J1818$-$1108} & &  \multicolumn{5}{c}{LS~5039} &&  \multicolumn{2}{c}{J1818$-$1108} && \multicolumn{5}{c}{J1825$-$1718}     \\
\cmidrule{2-3}     \cmidrule{5-9} \cmidrule{11-12}  \cmidrule{14-18}
         &  $\Delta_{\rm R.A.}$  & $\Delta_{\rm Dec.}$ && HPBW &   \multicolumn{2}{c}{$\Delta_{\rm R.A.}$}  & \multicolumn{2}{c}{$\Delta_{\rm Dec.}$} &&   $\Delta_{\rm R.A.}$  & $\Delta_{\rm Dec.}$  && HPBW &  \multicolumn{2}{c}{$\Delta_{\rm R.A.}$} &  \multicolumn{2}{c}{$\Delta_{\rm Dec.}$} \\
         &   [mas]  & [mas] && [mas$^{2}$ at $^{\circ}$] &   \multicolumn{2}{c}{[mas]}  & \multicolumn{2}{c}{[mas]} &&   [mas]  & [mas]  && [mas$^{2}$ at $^{\circ}$] &  \multicolumn{2}{c}{[mas]} &  \multicolumn{2}{c}{[mas]} \\
\midrule
A   &  $+$4.0 & $+$4.4  && 3.0$\times$1.2 at $-$4.3&   $+1.3$  & 0.5  & $+0.4$  & 0.7  &&  $+$1.1 & $+$0.3 && 6.8$\times$2.6 at \phantom{1}6.1   & $-$1.4 & 0.5 & $-$0.9  & 1.0 \\
B   &  $-$1.3 & $-$0.6  && 3.5$\times$1.3 at $-$0.9&   $-1.2$  & 0.3  & $+1.2$  & 0.4  &&  $+$0.7 & $-$1.0 && 7.2$\times$2.8 at \phantom{1}4.4   & $+$0.7 & 0.5 & $-$1.2  & 1.0 \\
C   &  $+$0.1 & $-$2.6  && 3.2$\times$1.3 at $-$2.8&   $-0.2$  & 0.4  & $-0.8$  & 0.5  &&  $+$0.2 & $+$1.2 && 6.8$\times$2.7 at \phantom{1}7.8   & $-$0.3 & 0.4 & $+$1.3  & 0.8 \\
D   &  $-$2.8 & $+$0.1  && 3.2$\times$1.3 at $-$1.9&   $+0.2$  & 0.5  & $-0.5$  & 0.6  &&  $+$0.2 & $+$1.5 && 6.8$\times$2.6 at \phantom{1}2.8   & $+$0.8 & 0.3 & $+$0.4  & 0.6 \\
E   &  $-$0.0 & $-$1.2  && 3.9$\times$1.3 at $-$5.4&   $-0.2$  & 0.5  & $-0.4$  & 0.8  &&  $-$2.2 & $-$2.0 && 7.1$\times$2.7 at 12.2             & $+$0.2 & 0.4 & $+$0.4  & 0.8 \\

\midrule
\end{tabular*}
\end{center}
\end{table*}

\subsection{Morphology} \label{morphology}

\begin{figure*}[] 
\center
\resizebox{1.0\hsize}{!}{\includegraphics[angle=0]{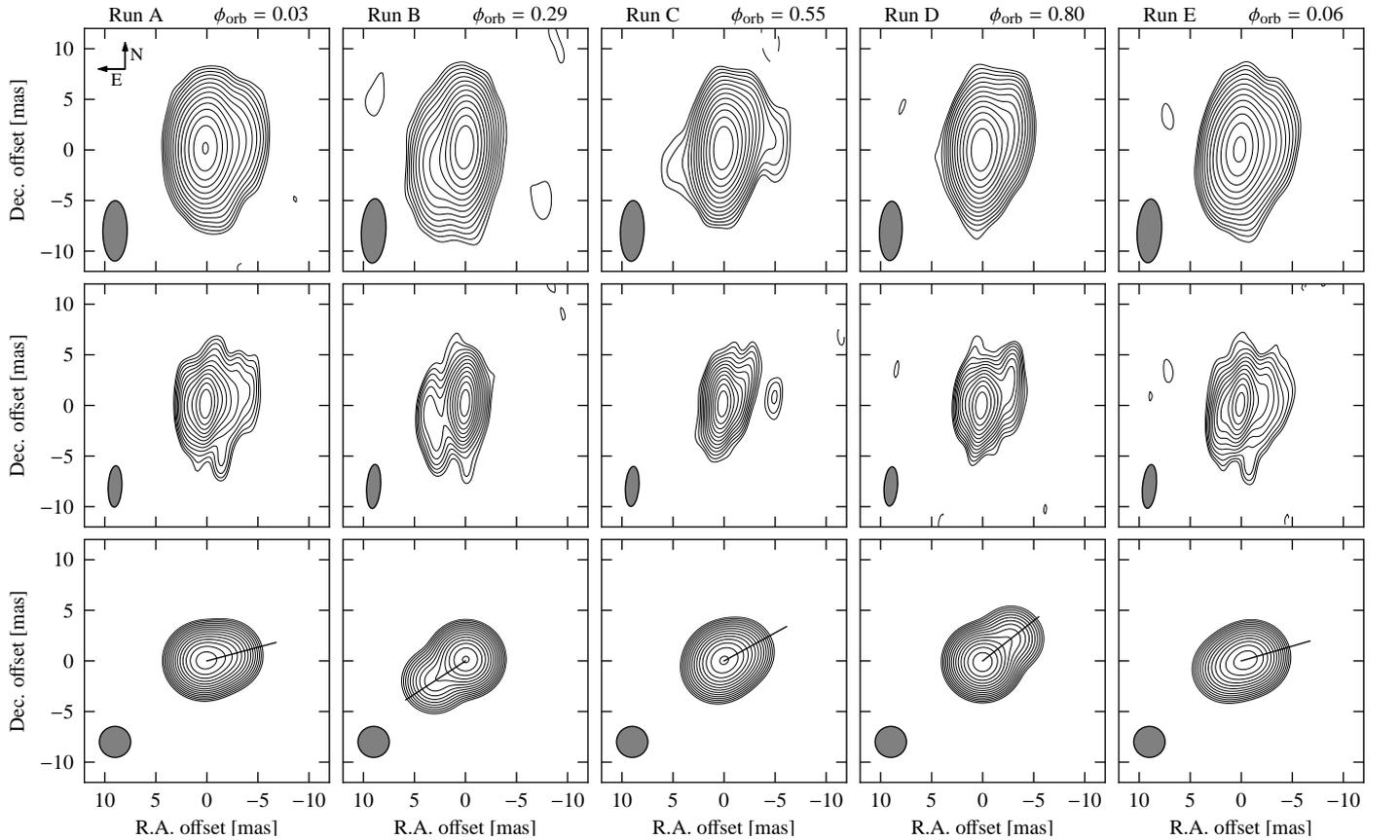}}
\caption{VLBA images of \object{LS~5039} at 5~GHz from project BR127, obtained during five consecutive days in July 2007. Each column, labelled with the run name and orbital phase, corresponds to one epoch (see Table~\ref{table:obsparam}). For each epoch, the first and second rows correspond to the self-calibrated images obtained with a natural and uniform weighting scheme, respectively. The third row shows the $uv$ components fitted to the data, convolved with a Gaussian circular beam with an area equal to the average synthesized beam of all images. The lines show the direction of the main extended component. The restoring beams are plotted in the bottom-left corner of each panel. Dashed contours are plotted at $-3$ times the rms noise of each image and solid contours start at 3 times the rms and scale with $2^{1/2}$. The Gaussian components fitted to the images are listed in Table~\ref{table:components}.}
\label{fig:br127}
\end{figure*}

The self-calibration of the data improves the S/N of the phase-referenced images while preserving the main features of the extended emission, partially avoiding the problems with the phase calibration caused by the ionosphere, at the expense of losing the astrometric information. In the first two rows of Fig.~\ref{fig:br127} we show the final self-calibrated images for each run, produced with natural and uniform weight, respectively. The five images with natural weight have synthesized beams with sizes about $6.1\times2.3$~mas$^{2}$ and P.A. of $-2^{\circ}$ and total flux densities of 27.9, 26.1, 24.4, 28.5, and 26.4~mJy, respectively, with a rms noise of the images of $\sim$0.06~mJy~beam$^{-1}$. On the other hand, the images with uniform weight have synthesized beams with sizes around $4.1\times1.3$~mas$^{2}$ and P.A. of $-4^{\circ}$. The total flux densities in the five images are 27.8, 26.7, 25.4, 28.5, and 27.0, with a rms noise of 0.08~mJy~beam$^{-1}$. The individual components fitted to the images with JMFIT are summarised in Table~\ref{table:components}, where we list the flux density, relative position, and size of each component. The last two columns show the deconvolved size of the components, which describes to a certain degree the intrinsic size of the component, deconvolved from the synthesized beam size. We also list the Gaussian components fitted to the $uv$ data (modelfit), which are not convolved with any beam. Most of the modelfit components are circular instead of elliptical. We plot these components in the last row of Fig.~\ref{fig:br127} convolved with an artificial circular beam with an area equal to the average synthesized beam of the rest of the images. As a guide, we plot a line towards the direction of the main extended component, which corresponds to P.A. of $-75$, 123, $-$61, $-$52, and $-74^{\circ}$, respectively.

\begin{table*}[t] 
\begin{center}
\caption{Gaussian components fitted to the images shown in Fig.~\ref{fig:br127}. The first two blocks show the fits to the images with natural and uniform weighting schemes, respectively, whereas the third block, labelled modelfit, shows fits to the $uv$ data. Each component is labelled with the corresponding letter of the epoch (A--E). Columns 2 and 3 list the peak and integrated flux density of each component. Columns 4--8 list the position offset and the size fitted to each component. The last two columns list the deconvolved size of the components fitted to the images. The modelfit block contains empty values because this fit has no dependence on a synthesized beam.}
\label{table:components}
\begin{tabular}{c r@{ $\pm$ }l r@{ $\pm$ }l r@{ $\pm$ }l r@{ $\pm$ }l r@{ $\pm$ }l r@{ $\pm$ }l r@{ $\pm$ }l r@{ $\times$ }l r }
\midrule
\midrule
Comp. & \multicolumn{2}{c}{$S_{\rm peak}$}   & \multicolumn{2}{c}{$S_{\rm tot}$} & \multicolumn{2}{c}{$\Delta_{\rm R.A.}$} & \multicolumn{2}{c}{$\Delta_{\rm Dec.}$} & \multicolumn{2}{c}{$\theta_{\rm maj}$} & \multicolumn{2}{c}{$\theta_{\rm min}$} & \multicolumn{2}{c}{$\theta_{\rm P.A.}$} & \multicolumn{3}{c}{Deconvolved size }  \\ 
     & \multicolumn{2}{c}{[mJy~b$^{-1}$]}& \multicolumn{2}{c}{[mJy]}         & \multicolumn{2}{c}{[mas]}               & \multicolumn{2}{c}{[mas]}               & \multicolumn{2}{c}{[mas]}              & \multicolumn{2}{c}{[mas]}              & \multicolumn{2}{c}{[$^{\circ}$]}        & \multicolumn{2}{c}{[mas$^{2}$]}  &\multicolumn{1}{c}{[$^{\circ}$]} \\
\midrule
\multicolumn{18}{c}{Natural} \\
\midrule
   A1 &  9.84 &  0.06 & 14.72 & 0.15 &    0.36 & 0.01  &    0.18 & 0.02 &  6.25 &  0.04 &  3.41 &  0.02 &  177.1 &    0.4 &  2.54 &  1.72 &  111.5  \\ 
   A2 &  6.27 &  0.07 &  6.27 & 0.07 &    0.00 & 0.01  &    0.00 & 0.03 &  5.97 &  0.01 &  2.39 &  0.01 &  179.6 &    0.1 & \multicolumn{3}{c}{Point-like}\\ 
   A3 &  3.04 &  0.06 &  7.02 & 0.2  & $-$2.05 & 0.04  &    0.17 & 0.06 &  7.52 &  0.15 &  4.37 &  0.09 &  169.5 &    1.4 &  4.91 &  3.19 &  145.5  \\ 
   B1 & 16.34 &  0.06 & 19.55 & 0.12 & $-$0.03 & 0.01  &    0.17 & 0.01 &  6.63 &  0.03 &  2.76 &  0.01 &  177.2 &    0.2 &  1.84 &  1.36 &    2.0  \\ 
   B2 &  2.99 &  0.06 &  4.04 & 0.13 &    3.05 & 0.02  & $-$1.64 & 0.07 &  7.64 &  0.16 &  2.71 &  0.06 &    1.0 &    0.7 &  4.27 &  1.03 &    9.2  \\ 
   C1 & 13.57 &  0.06 & 15.39 & 0.11 &    0.03 & 0.01  & $-$0.03 & 0.01 &  6.18 &  0.03 &  2.55 &  0.01 &  177.2 &    0.2 &  1.53 &  1.02 &  162.6  \\ 
   C2 &  3.30 &  0.06 &  7.18 & 0.17 & $-$0.75 & 0.04  &    0.81 & 0.05 &  6.84 &  0.12 &  4.41 &  0.08 &  161.1 &    1.5 &  4.58 &  1.99 &  119.1  \\ 
   D1 & 17.48 &  0.06 & 23.80 & 0.13 &    0.03 & 0.01  & $-$0.07 & 0.01 &  6.06 &  0.02 &  2.93 &  0.01 &  177.7 &    0.2 &  1.90 &  1.55 &   87.9  \\ 
   D2 &  2.95 &  0.06 &  3.27 & 0.12 & $-$2.86 & 0.02  &    2.06 & 0.06 &  6.19 &  0.13 &  2.33 &  0.05 &  179.9 &    0.8 &  2.09 &  0.33 &   16.3  \\ 
   E1 &  9.66 &  0.07 & 15.75 & 0.17 &    0.35 & 0.01  & $-$0.09 & 0.02 &  6.91 &  0.05 &  3.56 &  0.03 &  170.3 &    0.4 &  3.38 &  1.90 &  132.8  \\ 
   E2 &  5.70 &  0.07 &  5.70 & 0.07 &    0.00 & 0.01  &    0.00 & 0.03 &  6.30 &  0.01 &  2.39 &  0.01 &  176.8 &    0.1 & \multicolumn{3}{c}{Point-like}\\ 
   E3 &  2.41 &  0.07 &  3.94 & 0.17 & $-$2.58 & 0.04  & $-$0.08 & 0.09 &  7.45 &  0.21 &  3.31 &  0.10 &  168.0 &    1.3 &  4.30 &  1.61 &  148.1  \\ 
\midrule
\multicolumn{18}{c}{Uniform} \\
\midrule
   A1 &  7.45 &  0.08 &  7.45 & 0.08 &    0.00 & 0.01 &    0.00 & 0.02 &  4.09 &  0.00 &  1.34 &  0.00 &  177.5 &    0.1 & \multicolumn{3}{c}{Point-like}  \\ 
   A2 &  1.67 &  0.07 &  9.39 & 0.44 & $-$1.53 & 0.08 &    0.38 & 0.12 &  6.72 &  0.27 &  4.58 &  0.19 &    169 &      4 &  5.41 &  4.28 &  156.1  \\ 
   A3 &  5.43 &  0.07 & 11.71 & 0.21 &    0.41 & 0.02 &    0.11 & 0.02 &  4.10 &  0.05 &  2.88 &  0.04 &  168.8 &    1.5 &  2.63 &  0.00 &    7.8  \\ 
   B1 & 13.87 &  0.08 & 18.38 & 0.17 & $-$0.05 & 0.01 &    0.19 & 0.01 &  4.58 &  0.03 &  1.71 &  0.01 &  176.7 &    0.2 &  1.39 &  1.02 &   11.2  \\ 
   B2 &  2.04 &  0.08 &  5.00 & 0.26 &    2.92 & 0.04 & $-$1.62 & 0.11 &  6.43 &  0.25 &  2.25 &  0.09 &    1.6 &    1.3 &  4.76 &  1.71 &    6.5  \\ 
   C1 & 13.34 &  0.08 & 18.45 & 0.16 &    0.07 & 0.01 &    0.05 & 0.01 &  4.29 &  0.02 &  1.68 &  0.01 &  174.3 &    0.2 &  1.66 &  1.00 &  161.3  \\ 
   C2 &  2.26 &  0.08 &  3.12 & 0.16 & $-$1.86 & 0.02 &    1.08 & 0.07 &  5.16 &  0.17 &  1.40 &  0.05 &  169.7 &    0.8 &  3.36 &  0.00 &  161.9  \\ 
   D1 & 13.14 &  0.08 & 22.89 & 0.19 &    0.04 & 0.01 & $-$0.08 & 0.01 &  4.13 &  0.02 &  2.15 &  0.01 &  175.5 &    0.3 &  1.70 &  1.39 &   86.0  \\ 
   D2 &  2.32 &  0.08 &  4.04 & 0.19 & $-$2.73 & 0.02 &    1.90 & 0.07 &  5.17 &  0.17 &  1.71 &  0.06 &  176.5 &    1.0 &  3.42 &  1.10 &  177.7  \\ 
   E1 & 10.68 &  0.08 & 10.68 & 0.08 &    0.00 & 0.00 &    0.00 & 0.01 &  4.34 &  0.01 &  1.34 &  0.01 &  174.4 &    0.1 & \multicolumn{3}{c}{Point-like}  \\ 
   E2 &  2.65 &  0.07 &  10.1 &  0.3 & $-$1.33 & 0.05 &    0.29 & 0.07 &  5.70 &  0.16 &  3.90 &  0.11 &    169 &      3 &  3.88 &  3.46 &  128.8  \\ 
   E3 &  3.47 &  0.08 &  4.90 & 0.17 &    1.63 & 0.02 & $-$0.80 & 0.05 &  5.00 &  0.11 &  1.64 &  0.04 &  173.5 &    0.7 &  2.49 &  0.94 &  170.6  \\ 
\midrule
\multicolumn{18}{c}{Modelfit} \\
\midrule
   A1 & \multicolumn{2}{c}{---}   &  12.4 &  1.0 &     0.52 &  0.08 &    0.03 &  0.04 &   2.60 &  0.10 &            \multicolumn{4}{c}{Circular} &  \multicolumn{2}{c}{---}   &  ---    \\ 
   A2 & \multicolumn{2}{c}{---}   &   8.0 &  0.5 &     0.00 &  0.01 & $-$0.05 &  0.02 &   \multicolumn{2}{c}{0.0} & \multicolumn{4}{c}{Point-like}                                                  &  \multicolumn{2}{c}{---}   &  ---    \\ 
   A3 & \multicolumn{2}{c}{---}   &   7.4 &  1.0 &  $-$2.00 &  0.23 &    0.54 &  0.08 &   3.79 &  0.16 &            \multicolumn{4}{c}{Circular} &  \multicolumn{2}{c}{---}   &  ---    \\ 
   B1 & \multicolumn{2}{c}{---}   &  18.4 &  0.5 &  $-$0.05 &  0.01 &    0.16 &  0.01 &   1.11 &  0.01 &            \multicolumn{4}{c}{Circular} &  \multicolumn{2}{c}{---}   &  ---    \\ 
   B2 & \multicolumn{2}{c}{---}   &   4.5 &  0.5 &     3.04 &  0.02 & $-$2.00 &  0.07 &    4.2 &  0.3  &     1.2  &  0.1                &   14.6 &    2.1 &  \multicolumn{2}{c}{---}   &  ---    \\ 
   C1 & \multicolumn{2}{c}{---}   &  13.1 &  1.0 &  $-$0.02 &  0.01 &    0.19 &  0.02 &   0.57 &  0.06 &            \multicolumn{4}{c}{Circular} &  \multicolumn{2}{c}{---}   &  ---    \\ 
   C2 & \multicolumn{2}{c}{---}   &   3.7 &  0.5 &  $-$1.66 &  0.03 &    0.92 &  0.07 &   0.76 &  0.09 &            \multicolumn{4}{c}{Circular} &  \multicolumn{2}{c}{---}   &  ---    \\ 
   C3 & \multicolumn{2}{c}{---}   &   4.8 &  1.0 &     0.77 &  0.14 & $-$0.81 &  0.17 &   2.09 &  0.13 &            \multicolumn{4}{c}{Circular} &  \multicolumn{2}{c}{---}   &  ---    \\ 
   D1 & \multicolumn{2}{c}{---}   &  22.6 &  0.5 &     0.02 &  0.01 & $-$0.13 &  0.01 &   1.63 &  0.01 &            \multicolumn{4}{c}{Circular} &  \multicolumn{2}{c}{---}   &  ---    \\ 
   E2 & \multicolumn{2}{c}{---}   &   4.8 &  0.5 &  $-$2.77 &  0.04 &    2.20 &  0.08 &   1.64 &  0.09 &            \multicolumn{4}{c}{Circular} &  \multicolumn{2}{c}{---}   &  ---    \\ 
   E1 & \multicolumn{2}{c}{---}   &   5.3 &  1.0 &     1.52 &  0.06 & $-$0.84 &  0.09 &   1.57 &  0.13 &            \multicolumn{4}{c}{Circular} &  \multicolumn{2}{c}{---}   &  ---    \\ 
   E2 & \multicolumn{2}{c}{---}   &   9.3 &  0.5 &  $-$0.00 &  0.01 & $-$0.06 &  0.02 &   0.18 &  0.15 &            \multicolumn{4}{c}{Circular} &  \multicolumn{2}{c}{---}   &  ---    \\ 
   E3 & \multicolumn{2}{c}{---}   &  10.5 &  1.0 &  $-$1.23 &  0.13 &    0.36 &  0.05 &   3.66 &  0.16 &            \multicolumn{4}{c}{Circular} &  \multicolumn{2}{c}{---}   &  ---     \\ 
   
\midrule
\end{tabular}
\end{center}
\end{table*}

All images show a main core component and extended emission with changing P.A. During run~A, which was obtained soon after periastron (orbital phase $\sim0$), the source is extended westwards, although the core is better described with two components. This double core has an important contribution eastwards, at a distance of $\sim0.5$~mas. For the image obtained 24~h later, at orbital phase 0.29, a fast morphological change occurs and the main extended component is detected towards South-East. Run~C, obtained soon after apastron, displays bipolar extended emission. The emission towards North-West is in fact composed by two components, although only one component could be fitted. The faint component towards South-East could not be fitted to the data and therefore it is not listed in Table~\ref{table:components}. Run~D is well described with a main core and a bright component towards North-West. Finally, the images from run~E, also obtained short after the periastron passage, recover the same morphology of run~A. This shows that the morphology of \object{LS~5039} is periodic on consecutive orbital cycles.

We also divided each data set in two blocks to check for possible intraday variations. We obtained modelfit solutions and produced images for the divided $uv$ data sets. However, it was not possible to obtain reliable morphological differences because of the very different $uv$ coverage at the beginning and the end of each run.

\section{Compilation of VLBI observations} \label{compilation}

To try to confirm the repeatability of the changing morphological structure of \object{LS~5039} found in our five consecutive days campaign (project BR127), we compiled the available archival VLBI data of the source that provides an angular resolution similar to the one explored here. We compiled images from other eleven observations conducted at frequencies between 1.7 and 8.5~GHz between 1999 and 2009. A log of these observations is shown in Table~\ref{table:vlbi}. The orbital phases were computed with the ephemerides from \cite{casares12} and have uncertainties below 0.022.

\newcommand{\phm} {\phantom{$-$}}
\begin{table}[t!] 
\begin{center}
\caption{Log of the VLBI observations of \object{LS~5039}. Project codes starting with B correspond to VLBA observations, E to EVN observations, and G to Global (VLBA+EVN) observations.}
\label{table:vlbi}
\begin{tabular}{@{\extracolsep{-5pt}}@{}lccc r@{ at~~~}r   c@{}}
\midrule
\midrule
 Project   & Epoch   &  Orbital  &  Freq  & \multicolumn{2}{c}{HPBW}        & rms  \\
           & Y-M-D   &  phase    & [GHz]  & \multicolumn{2}{c}{[mas$^{2}$ at $^{\circ}$]}&  [mJy~b$^{-1}$]     \\
\midrule
BP051       & 1999-05-08   &  0.16& 5.0  &  $3.4\times1.2$ & \phm0.4  &  0.11   \\
\midrule
ER011       & 2000-03-01   &  0.39& 5.0  &  $7.6\times7.0$ &   $-$14  &  0.10   \\
\midrule
GR021A      & 2000-06-03   &  0.49& 5.0  &  $3.4\times1.2$ & \phm0.0  &  0.08   \\
GR021B      & 2000-06-08   &  0.77& 5.0  &  $3.4\times1.2$ & \phm0.0  &  0.11   \\
\midrule
BD087G      & 2004-06-11   &  0.58& 8.5  &  $2.6\times1.1$ &    12.2  &  0.11   \\
BD105A      & 2005-06-11   &  0.02& 8.4  &  $2.9\times1.3$ & \phm7.9  &  0.09   \\
BD105G      & 2006-01-29   &  0.51& 8.4  &  $3.4\times1.2$ & \phm7.3  &  0.10   \\
\midrule
EF018A      & 2007-03-01   &  0.89& 5.0  &  $5.9\times4.2$ & \phm5.3  &  0.14   \\
EF018B      & 2007-03-03   &  0.40& 5.0  &  $6.9\times4.1$ & \phm7.3  &  0.04   \\
EF018C      & 2007-03-05   &  0.91& 5.0  &  $4.5\times3.6$ & \phm8.1  &  0.11   \\
\midrule
BR127A      & 2007-07-05   &  0.03& 5.0  &  $6.0\times2.4$ & $-$0.4   &  0.06   \\
BR127B      & 2007-07-06   &  0.29& 5.0  &  $6.4\times2.4$ & $-$3.0   &  0.06   \\
BR127C      & 2007-07-07   &  0.55& 5.0  &  $6.0\times2.3$ & $-$2.2   &  0.06   \\
BR127D      & 2007-07-08   &  0.80& 5.0  &  $5.9\times2.2$ & $-$2.3   &  0.07   \\
BR127E      & 2007-07-09   &  0.06& 5.0  &  $6.3\times2.4$ & $-$3.2   &  0.07   \\
\midrule
EM074       & 2009-03-07   &  0.46& 1.7  &  $26\times5.8$ &  \phm7.3  &  0.04   \\
\midrule
\end{tabular}
\end{center}
\end{table}

There are three observations at 5~GHz (6~cm wavelength) from 1999 and 2000 that correspond to the projects BP051 and GR021A--B, which included the VLBA and the phased VLA array as an additional VLBI station. These results were already presented in \cite{paredes00} and \cite{ribo08}, where the interested reader can find detailed information on the data and the data reduction process.

Between 2004 and 2006, three runs were observed with the VLBA at 8.5~GHz (3.6~cm wavelength) as part of a long-term astrometric project (PI: V. Dhawan). The observational codes are BD087G, BD105A, and BD105G. The data were correlated with a lower sensitivity provided by 128~Mbps, and the final resolution is slightly better than for the other projects thanks to the higher frequency of the observations. We produced three self-calibrated images with natural weight. The obtained synthesized beams and noises are shown in Table~\ref{table:vlbi}. More details on the data reduction can be found in \cite{moldon12astrom}. The total flux densities recovered for BD087G, BD105A, and BD105G were 18.6, 17.6, 20.2~mJy, respectively.

We also present three self-calibrated images from project EF018 (PI: R. Fender), which consists of 3 runs (A--C) at 5.0~GHz obtained with the European VLBI Network (EVN) in 2007. The runs were separated by 2 days. The antennas used were Effelsberg, Westerbork, Jodrell Bank (Lovell), Onsala, Medicina, Noto, Torun, Nanshan (Urumqi), Sheshan (Shanghai), and Hartebeesthoek. This array provides good sensitivity, although each run lasted only 4~hours and the elevation of the source was lower than in the previous projects. The longest baselines in these observations were from the stations in Nanshan (Urumqi) and Hartebeesthoek, although the $uv$ coverage for those baselines is poor and the final resolution is lower. The total flux densities measured were 26.2, 19.1, and 27.4~mJy, respectively. The lower flux density and noise of epoch EF018B is probably an artifact of the amplitude self-calibration and not a real change in the source flux density.

\begin{figure*}[] 
\center
\resizebox{1.0\hsize}{!}{\includegraphics[angle=0]{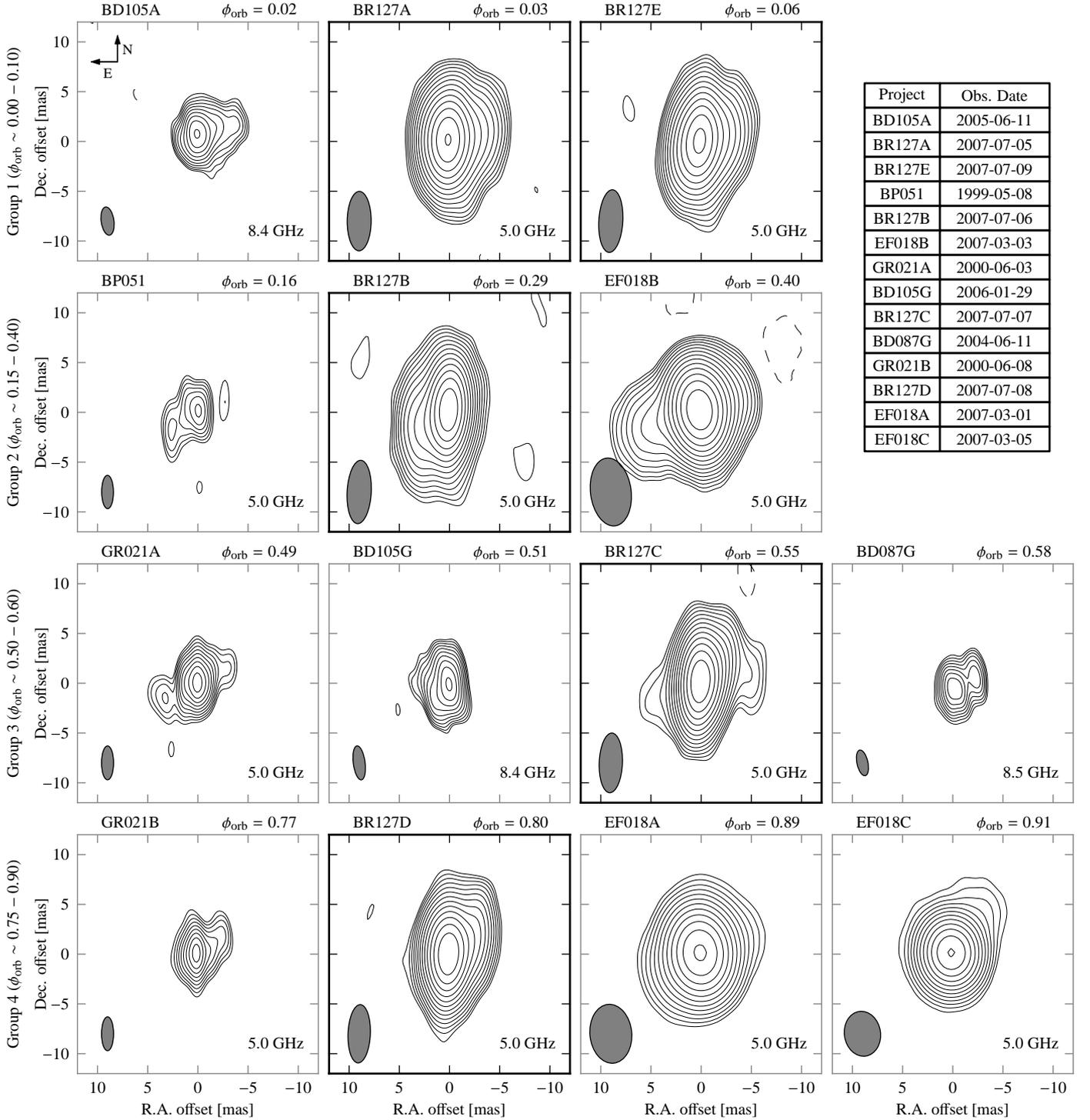}}
\caption{VLBI self-calibrated images of \object{LS~5039} at 5.0 and 8.5~GHz (see Table~\ref{table:vlbi} for details on the projects). The plotting parameters are the same as in Fig.~\ref{fig:br127}. The panel order is in increasing orbital phase from left to right and from top to bottom, and the images have been grouped according to similar morphological features. The project code and orbital phase are quoted on the top part of each panel. The images from project BR127, obtained during the same orbital cycle, have bold axes. Dashed contours are plotted at $-3$ times the rms noise of each image and solid contours start at 3 times the rms and scale with $2^{1/2}$, except for GR021A-B and BP051, which start at 5$\sigma$ as in the original publications. The rms noise of each image can be found in Table~\ref{table:vlbi}.}
\label{fig:vlbiall}
\end{figure*}

In Fig.~\ref{fig:vlbiall} we show the self-calibrated images of these projects. The images in the panels are ordered in increasing phase, from left to right and from top to bottom. We distributed them in four groups, which correspond to images with similar orbital phases and that show approximately a similar morphology. For a quick reference, the inset table included in Fig.~\ref{fig:vlbiall} shows the dates of the images in the panels. Although they were obtained with different instruments, frequencies, and at very different epochs, the images in each group share common morphological trends. Each group can be approximately characterised with the morphology described in Sect.~\ref{morphology} for the runs BR127A--D.

\subsection{Morphology at larger scales} \label{largerscales}

\begin{figure} 
\center
\resizebox{1.0\hsize}{!}{\includegraphics[angle=0]{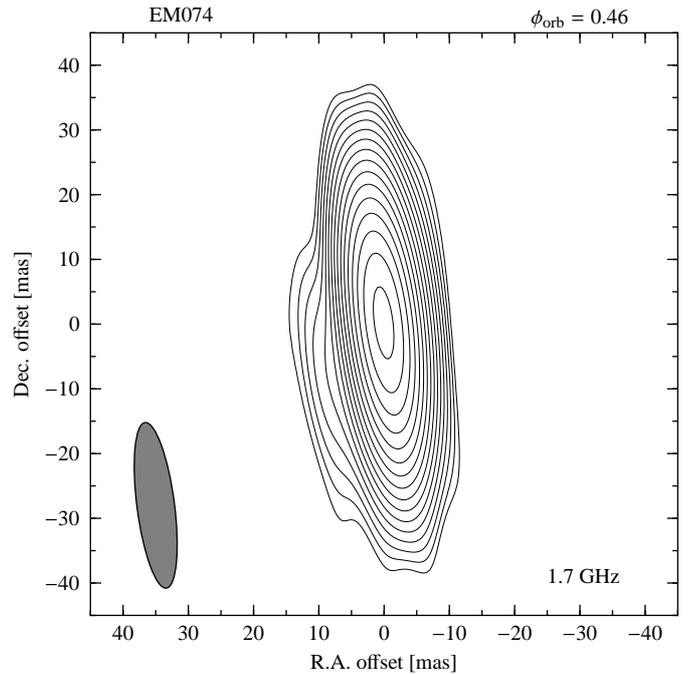}}
\caption{Self-calibrated image of \object{LS~5039} at 1.7~GHz obtained with the EVN in 2009. The plot parameters are the same as in Fig.~\ref{fig:vlbiall}, although a larger angular scale is shown here.}
\label{fig:em074}
\end{figure}

We observed \object{LS~5039} at 1.7~GHz (18~cm wavelength) with the EVN in 2009, project code EM074. This is a deep pointed observation on \object{LS~5039}, conducted without phase referencing. The total time on source was 5.8~h, and 9 stations participated: Effelsberg, Westerbork, Cambridge (32~m), Jodrell Bank (Lovell), Onsala, Medicina, Noto, Torun, and Urumqi. At this frequency the resolution is lower, although the long East-West baselines provide good resolution in right ascension. We produced a uniform weighting scheme image with a synthesized beam of $26\times5.8$~mas$^{2}$ at P.A. $7^{\circ}$ and a rms noise level of 0.04~mJy~beam$^{-1}$. The total flux density of the source was 33.5~mJy at 1.7~GHz. The self-calibrated image is shown in Fig.~\ref{fig:em074}. The source shows a morphology more similar to the images in group~2, although it was observed at orbital phase 0.46, which is closer to the ones of group~3. This can be explained by the longer lifetime of electrons emitting synchrotron radiation at lower frequencies.

On the other hand, the self-calibrated image from project ER011 provides lower angular resolution than the images in Fig.~\ref{fig:vlbiall} and it is not reproduced here because it is already published in \cite{paredes02}. The image was obtained at orbital phase 0.39. The innermost region of the core shows extended emission towards South-East, as it happens in the group~2 images in Fig.~\ref{fig:vlbiall}, although additional bipolar diffuse components extend up to $\sim20$~mas in opposite directions. \cite{paredes02} also present a lower resolution image obtained with MERLIN showing extended emission with P.A. of $150^{\circ}$ up to $170$~mas. Finally, we note that lower resolution VLA images show some hints of extended emission in the same North-West/South-East direction at angular distances of $\gtrsim0.1^{\prime\prime}$ \citep{marti98}. This confirms that \object{LS~5039} contains an additional extended and diffuse component that cannot be traced with the high-resolution images described in Sect.~\ref{morphology} and \ref{compilation}.

\section{Model} \label{model}

In this section we try to reproduce the changing radio morphology of \object{LS~5039} by modelling the emission produced by an outflow of electrons accelerated as a consequence of the interaction of the stellar wind and the wind of the putative young pulsar. We are interested in describing the orientation of the P.A. of the emission up to scales of $\sim5$--10~mas. The formation and evolution of the outflow is complex, and hydrodynamical and magnetohydrodynamical simulations are needed to describe in detail the source radio emission (see the discussion in Sect.~\ref{introduction}). Our aim here is to check if the general scenario of colliding winds is compatible with the main features and the continuous morphological changes found at scales up to 10~mas.

\begin{figure*}[ht!!!!!] 
\center
\resizebox{1.0\hsize}{!}{\includegraphics[angle=0]{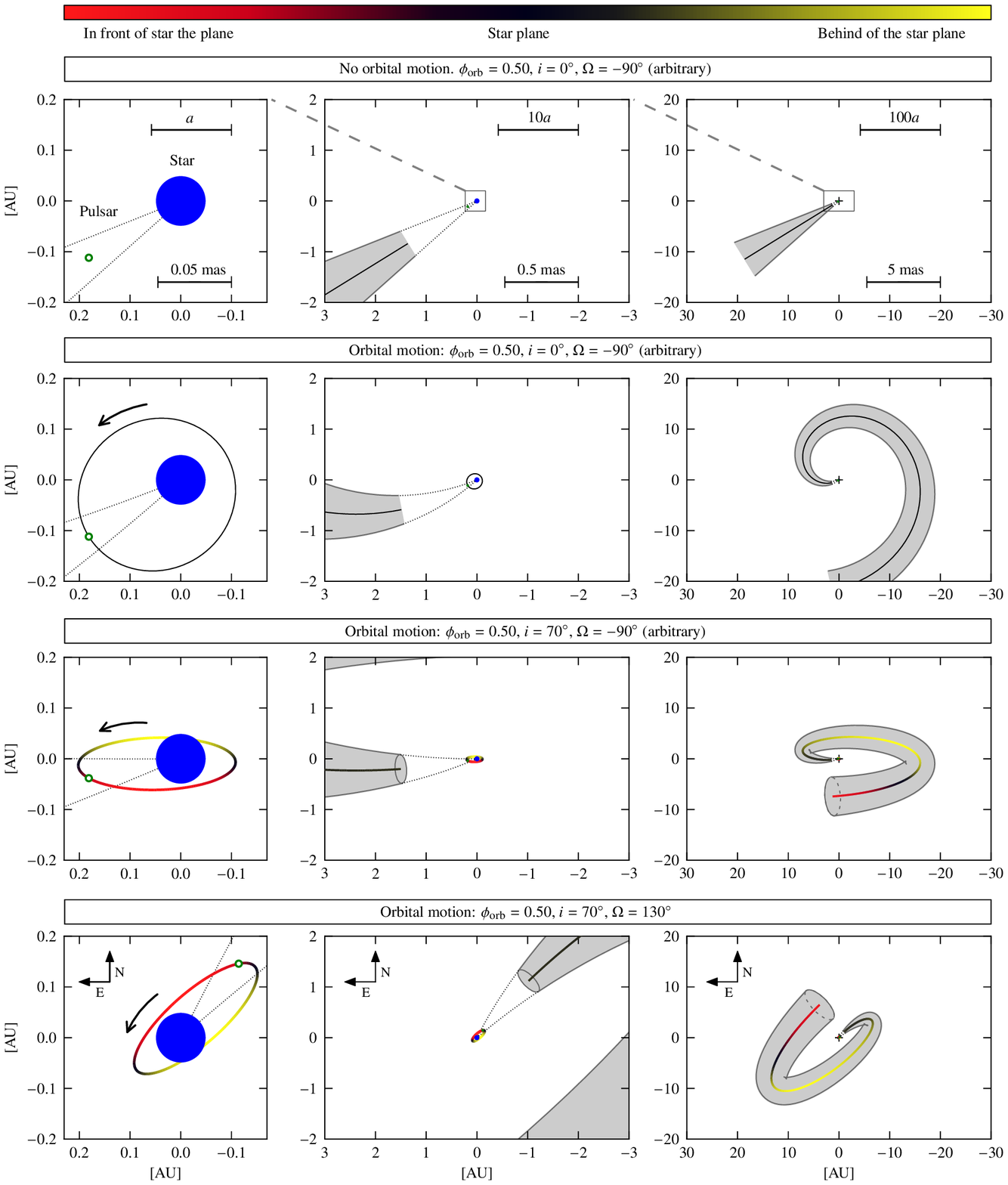}}
\caption{Sketch of the orbit orientation and the emission region of the model. From left to right the panels show three different scales: the orbit ($a$), the acceleration region ($10a$), and the whole flow ($100a$). The particle acceleration occurs at a distance of $10a$ from the massive star (blue circle) behind the pulsar (empty green circle) and the flow of emitting particles (grey area) expands forming  a cone while travelling away from the binary system. The star plane is the plane of the sky intersecting the star. The colours in the orbit and the flow indicate regions situated in front of the star plane (red, closer to the observer than the star) and behind the plane (yellow, farther than the star). The curved arrows indicate the orbital motion of the pulsar. The first row shows the general scenario without orbital motion, whereas the rest of the rows include the orbital motion and different projections of the orbit, indicated in the labels. The last row of panels shows the orbit projected with our best estimate of $i$ and $\Omega$.}
\label{fig:sketch}
\end{figure*}

We assume a mass for the companion star of 33~M$_{\odot}$ and an eccentricity of the system of 0.35 \citep{casares12}. The semimajor axis of the orbit for a 1.4~M$_{\odot}$ neutron star is approximately $a=2.4\times10^{12}$~cm, or 0.05~mas for a distance to the source of 2.9~kpc \citep{casares12}. The semimajor axis does not change significantly for compact object masses below 5~M$_{\odot}$. \cite{bosch-ramon12} identified regions where strong shocks are produced at distances between 5 and $10a$ from the massive star, towards the direction of the pulsar. Here we will assume that the extended radio emission is produced by electrons accelerated at a distance of 10 times the semimajor axis of the orbit, $\sim$0.5~mas at 2.9~kpc, and neglect the high-energy processes taking place at shorter distances. We show a sketch of the scenario in the first row of panels of Fig.~\ref{fig:sketch}, where the grey area indicates the flow of particles accelerated at a distance of $10a$ from the star and are travelling away from it. We consider that the injected electrons follow an energy power law-distribution, $N_{e}\propto\gamma_{e}^{-p}$. The global radio spectral index of the source is $-0.46$ \citep{marti98}, and therefore we use an electron index $p=2$. For simplicity, we assume a constant injection rate of $Q_{\rm inj} = 10^{35}$~erg~s$^{-1}$ (see below) along the orbit, for electron energies of $1<\gamma_{e}<10^{4}$. Higher electron energies do not significantly contribute to the radio emission at the considered frequencies. On the other hand, a range for the lower electron energy of $\sim1-100$ slightly modifies the most inner part of the emission, but does not significantly affect the larger source structure at $2-5$~mas. The model is not detailed enough to constrain this value and therefore we assume the minimum energy possible. We note that much higher minimum energies are expected in the primary shock between the stellar and the pulsar wind \citep{kennel84} but we are here considering a secondary shock at larger distances. \cite{zabalza11} inferred a spin-down luminosity for the putative pulsar in \object{LS~5039} of $L_{\rm sd}\sim(3$--$6)\times10^{36}$~erg~s$^{-1}$, and therefore our injected energy accounts for $\sim2$\% of this pulsar spin-down luminosity. The energy spectrum was divided in 200 logarithmic intervals for the computations.

For distances from the massive star larger than $10a$, the evolution of the electron distribution is determined by the radiative losses (synchrotron and inverse Compton in the Thomson regime) and adiabatic cooling. We assume that the particles travel away from the massive star forming an outflow with conical shape (see Fig.~\ref{fig:sketch}). This is conical shape is a coarse estimation, as the ambient pressure and the hydrodynamical instabilities severely modify the flow structure \citep{bosch-ramon12}. We use an opening angle for the cone of $20^{\circ}$, of the same order as the post-shock flow structure seen in \citet{bosch-ramon12} (see also \citealt{bogovalov08}), although the final flux density distribution is not very sensitive to this value. We note that this angle is much lower than the asymptotic opening angle of the contact discontinuity, which for this system is about $\sim75^{\circ}$ assuming a pulsar with spin-down luminosity of $5\times10^{36}$~erg~s$^{-1}$, with a stellar wind with a mass loss rate of $2.5\times10^{-7}$~M$_{\odot}$~yr$^{-1}$ and wind velocity at infinity of $2.4\times10^{3}$~km~s$^{-1}$ \citep{mcswain11}. We compute the electron energy distribution density $n(r, \gamma)$, where $r$ is the linear distance from the massive star, taking into account the conservation of the number of particles along the outflow through the continuity equation $N(r, \gamma)d\gamma=N(r_{0}, \gamma_{0})d\gamma_{0}$.

The outflow is bent as the particles travel away from the system because of the orbital motion of the pulsar (see second row of panels of Fig.~\ref{fig:sketch}). The emitting region is bent in the opposite direction of the orbital motion of the pulsar and it forms a spiral (see \citealt{dubus06} for a general description). The flow, similar to a cometary tail, is a projection of the orbit assuming a ballistic motion. However, the outflow, and in general the whole interaction structure, in fact should not follow a ballistic motion because of the effect of the Coriolis force and consequently the resulting spiral should be less open than for simple ballistic motion \citep{bosch-ramon11, bosch-ramon12}. Nevertheless, we approximate the outflow shape assuming a constant advection velocity of $0.04c$, which traces an structure similar but more open than the one found in the hydrodynamical simulations by \cite{bosch-ramon12}, because we find a better match with the images in this case. We note that those simulations were computed for a slightly different system and using a circular orbit.

We compute the electron density along the flow for $r$ between $10a$ and $140a$, although the farthest part ($\gtrsim100a$) does not contribute significantly to the emission. We divide the flow in 200 steps in $r$. The sizes of the steps is determined by the displacement of the pulsar along the orbit and are computed at constant increments of the true anomaly of the pulsar. Therefore, steps close to periastron are shorter, corresponding to time intervals of about 500~s each, whereas steps close to apastron are longer, corresponding to time intervals of about 3000~s each. These intervals are always at least one order of magnitude shorter than the fastest cooling time at each $r$ and $\gamma_{e}$. For the lower energy electrons, adiabatic cooling dominates along the flow, whereas for more energetic electrons, inverse Compton losses dominate up to distances of $\lesssim4$~mas. We compute the emissivity of each section of the flow assuming a magnetic field of $B= 1.0\left[\frac{2.4\times10^{12}{\rm cm}}{r} \right]$~G, which corresponds to $B\sim0.1$~G in the acceleration region. We note that this magnetic field is similar to the equipartition magnetic field inferred for non-thermal particles from VLBI observations, which is $\sim0.2$~G \citep{paredes00}.

The observed outflow geometry strongly depends on the orientation of the orbit on the sky, which is basically determined by two angles: the longitude of the ascending node, $\Omega$, measured from North to the East, and  the inclination of the orbit, $i$, which is a rotation with respect to the node axis (the orbital elements are defined in \citealt{smart30}). In Fig.~\ref{fig:sketch} we show different projections of these two angles at orbital phase 0.5. The images in Fig.~\ref{fig:vlbiall} show a privileged North-East/South-West direction, in a P.A. of $\sim130^{\circ}$. This constant direction for most part of the orbital phases suggests a high inclination of the orbit, already discussed in \cite{ribo08}. The orbital elements determine the flow orientation and the final flux density distribution. We searched for the combination of $\Omega$ and $i$ that better describes the source structure seen in Fig.~\ref{fig:vlbiall}. We used the electron density and the system geometry to compute the synchrotron emissivity of the electrons and project it on the sky. For every orbital phase in Fig.~\ref{fig:vlbiall}, we computed the sky flux density distribution at the corresponding frequency and convolved it with a beam equal to the corresponding synthesized beam of the image. The model accounts for part of the core component and the extended emission. We produced synthetic images in steps of 5$^{\circ}$ in $i$ and $\Omega$ independently and we found that the images in Fig.~\ref{fig:vlbiall} are better reproduced by the model for an inclination of $i\sim70^{\circ}$ and $\Omega\sim130^{\circ}$. Based on visual inspection, a reasonable range for these parameters is $i=60-75^{\circ}$ and $\Omega=120-140^{\circ}$. The rotation of the pulsar around the star is counter-clockwise, and therefore $i < 90^{\circ}$ (see \citealt{smart30}). Soon after the periastron passage the pulsar transits behind the star plane and after apastron it is in front of the star plane. For the inclination obtained the pulsar position is eclipsed during its superior conjunction.

\begin{figure*}[t] 
\center
\resizebox{1.0\hsize}{!}{\includegraphics[angle=0]{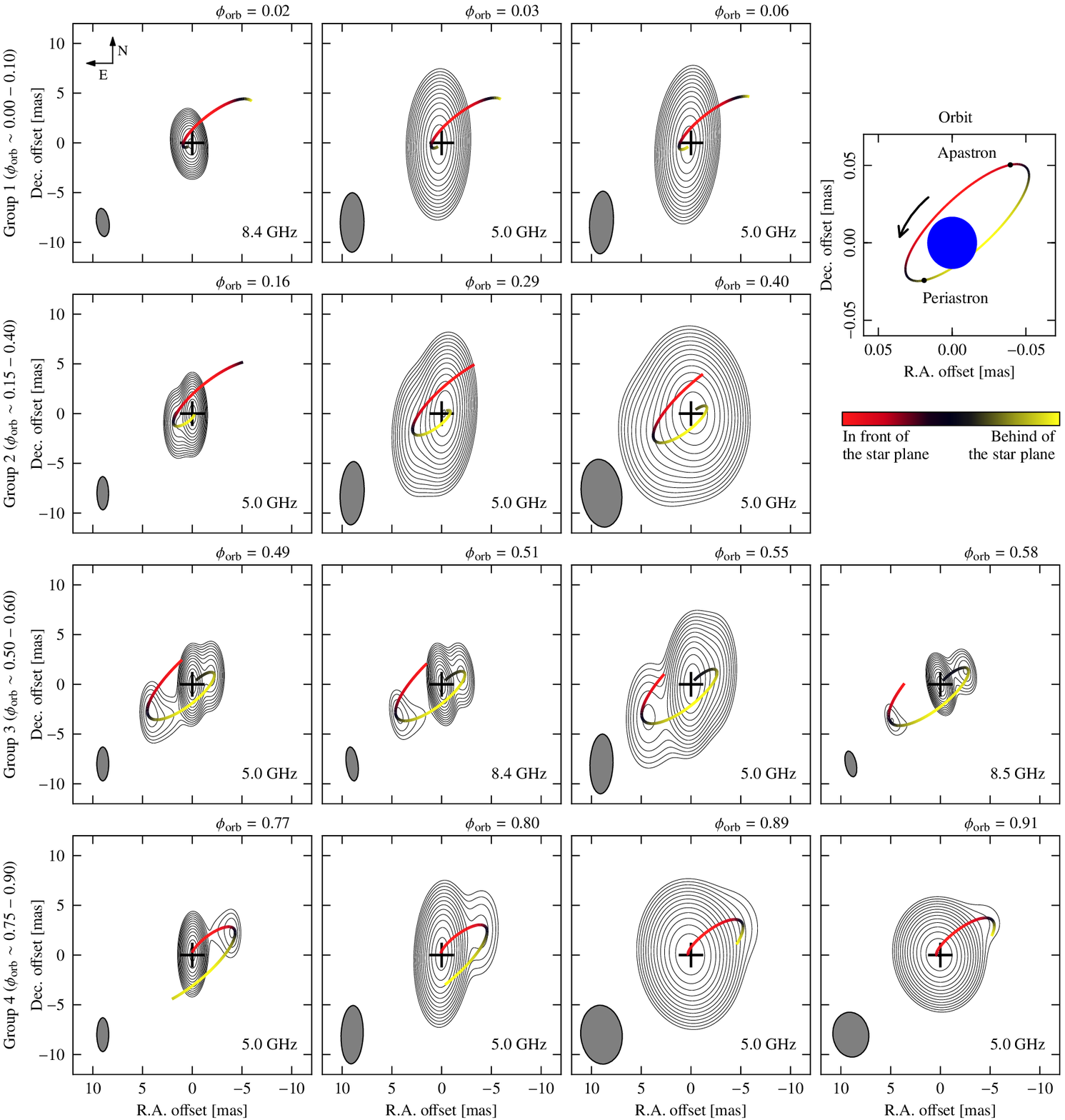}}
\caption{Synthetic images produced with our model, computed for the corresponding orbital phases of the different VLBI projects and convolved with the corresponding beam shown in Fig.~\ref{fig:vlbiall}. The black line traces the projected position of the flow axis of cooling particles, which is computed for linear distances from the star between 1.6 and 22~AU. The crosses mark the position of the main star, set at the origin. The panel in the top-right corner shows the system orbit, projected with an inclination of $70^{\circ}$ and a longitude of the ascending node of $130^{\circ}$.}
\label{fig:model}
\end{figure*}

The resulting synthetic images and the projected orbit are shown in Fig.~\ref{fig:model}. The first contour level for all plots is set to 0.06~mJy~beam$^{-1}$. The model produces total flux densities of around 10--16~mJy, which accounts for the extended flux density recovered with VLBI. However, this model does not account for the emission produced at distances below $\lesssim0.5$~mas, so we artificially included an additional 10~mJy point-like source at the position of the binary system to visually help to compare the results with the images in Fig.~\ref{fig:vlbiall}. This component approximately accounts for the peak flux density of the core component of the images. With this ad-hoc component, we recover a total flux density of the order of 15--18~mJy at 5~GHz and 13--16~mJy at 8.5~GHz, for the considered orbital phases, which are slightly below the measured values.

This model describes correctly the main features in most of the observed images, except for the observations shortly after the periastron passage. For that case (group~1), our model only accounts for the secondary component eastwards, but not for the main extended component towards West.The rest of the images are well described by this simple model. It produces a bright extended component for orbital phases 0.15--0.40 (group~2) and continuously develops extended emission towards North-West between phases 0.5 and 0.6 (group~3), while preserving the bipolar structure. The extended emission is then dominated by the North-West component until the next periastron passage. We note that this model predicts bipolar extended emission at certain orbital phases, although the outflow is not bipolar itself. In particular, the projected flow at phase 0.5, shown in the bottom-right panel of Fig.~\ref{fig:sketch}, accounts for the bipolar flux density distribution seen for instance in the image of project GR021A (see Figs.~\ref{fig:vlbiall} and \ref{fig:model}). At the same time, this approach justifies that the bipolar emission is not completely symmetric and that the two components show different P.A. \citep{ribo08}.

The differences between the model (Fig.~\ref{fig:model}) and the real data (Fig.~\ref{fig:vlbiall}), in particular the missing component in the images from group~1, can be explained by the oversimplification of the flow structure and the constant physical conditions at all orbital phases. For instance, we assumed that the acceleration region was at a constant distance from the massive star and that the energy injection was constant, although it is expected that these parameters depend on the orbital phase in an eccentric binary system. Therefore, additional electron populations, as well as more realistic time-dependent physical conditions should be included to produce this additional component. The model should also include considerations on the relativistic Doppler boosting possibly affecting the synchrotron emission \citep{dubus10}. However, including additional particle populations and assuming time-dependent physical conditions would require the addition of several free parameters, which is outside of the scope of this paper. Therefore, we do not try to include a complete flux density analysis or to predict flux density distributions. Also, predictions of displacements of the peak of the emission, such as the hint found in Sect.~\ref{astrometry}, cannot be obtained by this model, as these are produced at smaller distances. Despite this caveats, this model traces, except for the phases close to the periastron passage, all the main features of the VLBI images and the morphological variability measured in multifrequency and multiepoch observations, even when assuming simple and constant physical conditions.

\section{Discussion and conclusions} \label{conclusions}

VLBA observations at 5~GHz during five consecutive days show that \object{LS~5039} displays orbital morphological variability, showing one sided and bipolar structures, but recovering the same morphology when observing at the same orbital phase. The P.A.\ of the extended emission with respect to the core changes significantly on timescales of one day. Although the total flux density remains approximately constant during the orbital cycle, the flux density of the different components varies. Remarkably, the P.A. of the extended emission changes by $\sim160^{\circ}$ between phases 0.03 and 0.29 (24~h). The sensitivity of the observations is not enough to study changes on shorter timescales, within a few hours. We measured a displacement of the peak of the emission between orbital phases 0.29 and 0.55 of $2.3\pm0.6$~mas in P.A.\ $154^{\circ}$, nearly in the same direction of the extended emission. However, we consider this measurement as a hint of peak displacement because, given the limitations of the astrometry of these observations, we would require a 5-$\sigma$ significance to consider it a significant displacement. \cite{dubus06} predicted a continuous peak displacement forming an ellipse of a few mas along the orbit. The peak of the emission of the gamma-ray binary \object{LS~I~+61~303} at 8.4~GHz traces an ellipse of semimajor axis 1.5~AU, about four times the binary semimajor axis \citep{dhawan06}. Unfortunately our astrometric accuracy is not enough to measure these displacements. The source morphology during two consecutive periastron passages (run~A and E) is very similar, showing that the changes within the same orbital cycle are periodic. An interesting future project would be aimed to obtain accurate astrometry to explore the displacements of the peak of the emission, in particular at different frequencies, and also to measure variations of the source structure on timescales of a few hours.

We analysed the available VLBI observations of \object{LS~5039} in a consistent way. Images from data obtained between 1999 and 2009 at frequencies between 1.7 and 8.5~GHz show that the morphology at similar orbital phases is similar. When ordered in increasing phase, the images show an approximately continuous change. The observational conclusion of our VLBI analysis is that the morphological changes of \object{LS~5039} are periodic and show orbital modulation stable over several years. The gamma-ray binary \object{LS~I~+61~303} also shows a similar behaviour \citep{dhawan06}. Therefore, all gamma-ray binaries are expected to display periodic orbital modulation of their VLBI structure.

A simple model of an expanding outflow of relativistic electrons accelerated at a distance of $\sim10a$ and travelling away from the system can account for the main changes seen in the extended emission of \object{LS~5039}. The changes are explained for an inclination of the orbit $i\sim70^{\circ}$ and a longitude of the ascending node of $\Omega\sim130^{\circ}$. In the last row of panels in Fig.~\ref{fig:sketch} we show the orientation of the orbit on the sky projected with these parameters. The inclination of the orbit has deep implications in the physical properties of the binary system, because the orbital parameters restrict the mass of the compact object within the system. In particular, assuming a mass function of the binary system of 0.0045(6)~M$_{\odot}$ \citep{casares12}, a mass of the star of 33~M$_{\odot}$ implies that the mass of the compact object is 1.8--2.0~M$_{\odot}$, for an inclination between 75 and 60$^{\circ}$. However, the mass of the main star is barely constrained between 20 and 50~M$_{\odot}$, and consequently the inclinations inferred from the model are compatible with masses between 1.3--1.5 and 2.4--2.7~M$_{\odot}$, for the lower and the higher stellar mass limits, respectively. The uncertainty in the mass function adds an additional uncertainty of $\sim0.1$~M$_{\odot}$. The values obtained are in any case below 3~M$_{\odot}$, and therefore are compatible with the presence of a neutron star in the system. However, we note that this was one hypothesis of the model and therefore cannot be a proof. After applying a simple model, we conclude that that the presence of a young non-accreting pulsar in \object{LS~5039} agrees with the observed source structure at different orbital phases and produces a coherent scenario. However, this model does not account for all the emission produced at shorter distances and also fails to explain the main component of the extended emission seen shortly after the periastron passage. This is possibly due to unaccounted variability in the injection rate, geometry and localisation of the acceleration zone, and flow velocity of the electrons.

For gamma-ray binaries, the nature of the compact object, either accreting or not, is a fundamental ingredient for any model aimed to understand these peculiar systems. The most straightforward indication would be the detection of pulsations. However, direct detection may be unfeasible in binary systems with close orbits and powerful massive companions because of free-free absorption \citep{dubus06}. In addition, the mass function of the known gamma-ray binaries does not allow to discriminate between black holes or neutron stars. Therefore, an ongoing debate is still open regarding the nature of the compact object of gamma-ray binaries. We have presented high resolution radio images of \object{LS~5039} with which the scenarios can be tested. A first approach has shown that the source radio morphology can be explained with the presence of a young non-accreting pulsar. \object{PSR~B1259$-$63}, the only gamma-ray binary with a confirmed pulsar, shows variable extended radio structures \citep{moldon11_psr}, and \object{LS~I~+61~303} shows variable and periodic morphological changes \citep{dhawan06}. The common features among these three systems suggest that the known gamma-ray binaries contain young non-accreting pulsars.

\begin{acknowledgements}

We want to thank V.~Bosch-Ramon for very useful discussions on theoretical modelling. We are grateful to J.~Casares for allowing us to mention several new parameters of LS~5039 prior to publication. We also thank V. Zabalza and K. Iwasawa for helpful comments and discussions.
We thank the anonymous referee for very useful comments.
The Very Long Baseline Array is operated by the USA National Radio Astronomy Observatory, which is a facility of the USA National Science Foundation operated under co-operative agreement by Associated Universities, Inc.
The European VLBI Network (http://www.evlbi.org/) is a joint facility of European, Chinese, South African, and other radio astronomy institutes funded by their national research councils.
This research has made use of SAO/NASA's Astrophysics Data System.
We acknowledge support by the Spanish Ministerio de Ciencia e Innovaci\'on (MICINN) under grants AYA2010-21782-C03-01 and FPA2010-22056-C06-02.
J.M. acknowledges support by MICINN under grant BES-2008-004564.
M.R. acknowledges financial support from MICINN and European Social Funds through a \emph{Ram\'on y Cajal} fellowship.
J.M.P. acknowledges financial support from ICREA Academia.

\end{acknowledgements}

\bibliographystyle{aa}
\bibliography{biblio}

\end{document}